# A Conversation with Peter Huber

**Andreas Buja and Hans R. Künsch**

*Abstract.* Peter J. Huber was born on March 25, 1934, in Wohlen, a small town in the Swiss countryside. He obtained a diploma in mathematics in 1958 and a Ph.D. in mathematics in 1961, both from ETH Zurich. His thesis was in pure mathematics, but he then decided to go into statistics. He spent 1961–1963 as a postdoc at the statistics department in Berkeley where he wrote his first and most famous paper on robust statistics, "Robust Estimation of a Location Parameter." After a position as a visiting professor at Cornell University, he became a full professor at ETH Zurich. He worked at ETH until 1978, interspersed by visiting positions at Cornell, Yale, Princeton and Harvard. After leaving ETH, he held professor positions at Harvard University 1978–1988, at MIT 1988–1992, and finally at the University of Bayreuth from 1992 until his retirement in 1999. He now lives in Klosters, a village in the Grisons in the Swiss Alps.

Peter Huber has published four books and over 70 papers on statistics and data analysis. In addition, he has written more than a dozen papers and two books on Babylonian mathematics, astronomy and history. In 1972, he delivered the Wald lectures. He is a fellow of the IMS, of the American Association for the Advancement of Science, and of the American Academy of Arts and Sciences. In 1988 he received a Humboldt Award and in 1994 an honorary doctorate from the University of Neuchâtel. In addition to his fundamental results in robust statistics, Peter Huber made important contributions to computational statistics, strategies in data analysis, and applications of statistics in fields such as crystallography, EEGs, and human growth curves.

This conversation took place at Professor Huber's home in Klosters, Switzerland, on November 10, 2005.

*Andreas Buja is is Liem Sioe Liong/First Pacific Company Professor of Statistics, Statistics Department, The Wharton School, University of Pennsylvania, Philadelphia, Pennsylvania, USA (e-mail: buja.at.wharton@gmail.com). Hans R. Künsch is Professor and Chair, Department of Mathematics, ETH, CH-8092 Zürich, Switzerland (e-mail: kuensch@stat.math.ethz.ch).*

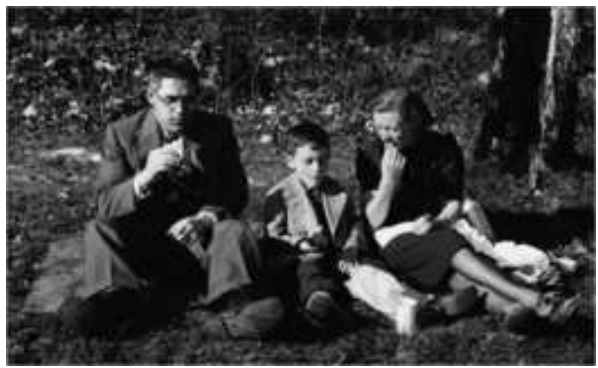

FIG. 1. *Peter with his parents, 1940.*

## STUDY YEARS AND THE MOVE INTO STATISTICS

**HK:** How did you find your way into the field of statistics?

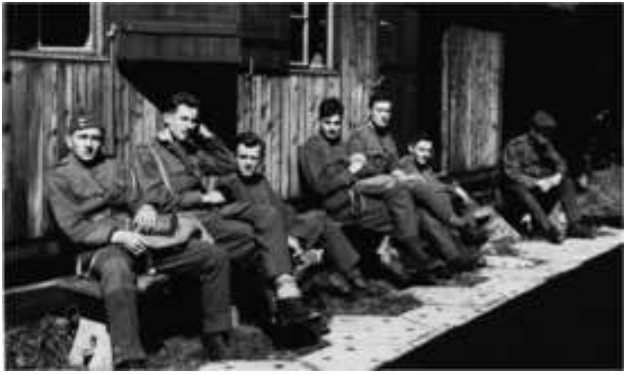

Fig. 2. *In the military, 1954. Peter is second from right.*

**PH:** More or less by accident. I started my career in pure mathematics, more specifically, in category theory. I noted a tendency in my research to move from the concrete to the abstract, but if I was already in category theory, and if I pushed further toward the abstract, I thought I would end up in empty space. This didn't seem like the right way to start a career. I wanted to start somewhere else, in a more concrete place in mathematics. I entertained a move to functional analysis, but around that time it so happened that ETH[1] was looking for a senior statistician. In its search ETH had contacted a couple of seniors, among them Erich Lehmann, but they had declined. So ETH decided to nurture local talent for statistics, but the question was who. In late 1960 I was approached by two professors at ETH, Walter Saxer and Eduard Stiefel. Saxer may have contacted me first, but it was Stiefel who had made the suggestion to try me. I thought hard about the proposal and ultimately concluded that mathematical statistics is not that far from functional analysis. The proposal started to make sense to me, and I wanted to give it a try. So I started to look into statistics.

At this point, the story merges with another, older story. At the time I was collaborating with B. L. van der Waerden, the famous algebraist at the University of Zurich next door to ETH, on writing a book on ancient astronomy. I had read van der Waerden's book on statistics, but I had never talked to him about the subject. Neither had I taken any courses in probability or statistics. I had sampled two courses by Saxer and by Linder, but I didn't last long because they were too low-level and, to be honest, not very captivating. Then I read van der Waerden's statistics text again, this time more carefully, and I read some of the books he recommended in his foreword, namely, Wald's "*Statistical decision functions*" and Doob's "*Stochastic processes.*" With this foundation it became clear that I should go to Berkeley to learn statistics. Beno Eckmann, my Ph.D. advisor and the leading algebraist and topologist at ETH at the time, tried to dissuade me; he wanted me to stay in topology. I couldn't see much of a risk, though, because if I didn't like statistics I could just change floors and spend my time in Berkeley's math department. So this is the circuitous story of how I got into statistics.

**AB:** Maybe you could tell us what ETH and the University of Zurich were like at the time, and what studying in these places was like. You spent a semester at the University and then switched to ETH.

**PH:** It was not easy to decide in which place to study, University of Zurich or ETH. The difference was that at the University you were essentially free to choose what to do and how to do it. You could study without a single exam up to the Ph.D. defense.

**HK:** That's surprising.

**PH:** A negative point was that you were never sure whether mandatory courses were offered. You couldn't be sure when you would finish your studies, without feedback from exams and with no disincentive for procrastination. ETH as a technical university was much more structured. You had a fixed course of study, and you could finish with a diploma (equivalent to a masters degree) after four years. ETH was on a yearly schedule and would start in fall, whereas the University was more loosely organized, and you could start also in spring. After high school and before the obligatory Swiss military service of 17 weeks, I had to make use of my time between graduation ("Maturität") in March and the beginning of boot camp in July. So in spring 1954 I enrolled at the University and sampled different courses. In particular, I attended van der Waerden's course on algebra. I found out that he was interested in the history of mathematics, and I was interested in Assyriology, Babylonian mathematics and the like.

**AB:** That interest goes back to high school...

**PH:** Yes, that's another story we may get to... In addition to learning algebra from van der Waerden,

---

[1] The "Swiss Federal Institute of Technology," abbreviated from German "Eidgenössische Technische Hochschule." At the time it was located in Zurich only. In 1969, a sister school was founded in Lausanne, abbreviated EPFL from French "École Polytechnique Fédéral, Lausanne."

I soon was in close contact with him working on the history of mathematics and astronomy. I liked his style, which was very direct. Looking back, I think, he had a decisive influence on me because he started me on writing papers and publishing them, and that helped.

**HK:** How many students were there in mathematics at that time?

**PH:** I don't remember the precise numbers, but at the University, in a graduate course such as algebra there were eight or ten students, and at ETH, when I started, the cohort of mathematics and physics together consisted of about 40, of which three quarters may have been physicists and one quarter mathematicians. The first two years mathematicians and physicists were essentially together all the time. The difference was that the mathematicians took astronomy in the first term while the physicists took chemistry, and if you were not sure, you took both, which I did. I wasn't sure which way I would go.

**HK:** Going back to the books you read when you got into statistics, you didn't mention Cramér. This would have been another plausible book that would have been around at that time.

**PH:** Yes, it was around. I may have looked into it, but I don't think I really read it. I liked van der Waerden's style, so I looked merely into the references he recommended, and among those he recommended were—I think he did have Cramér too—but let me have a look [grabs van der Waerden's book from the shelf and cites]: "It makes no sense to redevelop theories that are comprehensively treated by Kolmogorov, Carathéodory and Cramér." Subsequently he recommends Wald's "Sequential Analysis," his "Statistical decision functions," and Doob's "Stochastic processes." But I really did this only when I pondered the question of whether to go into statistics or not. And so I looked into statistics to find out (a) whether I would find it sufficiently attractive in the long run, and (b) whether I could get a feel for the subject.

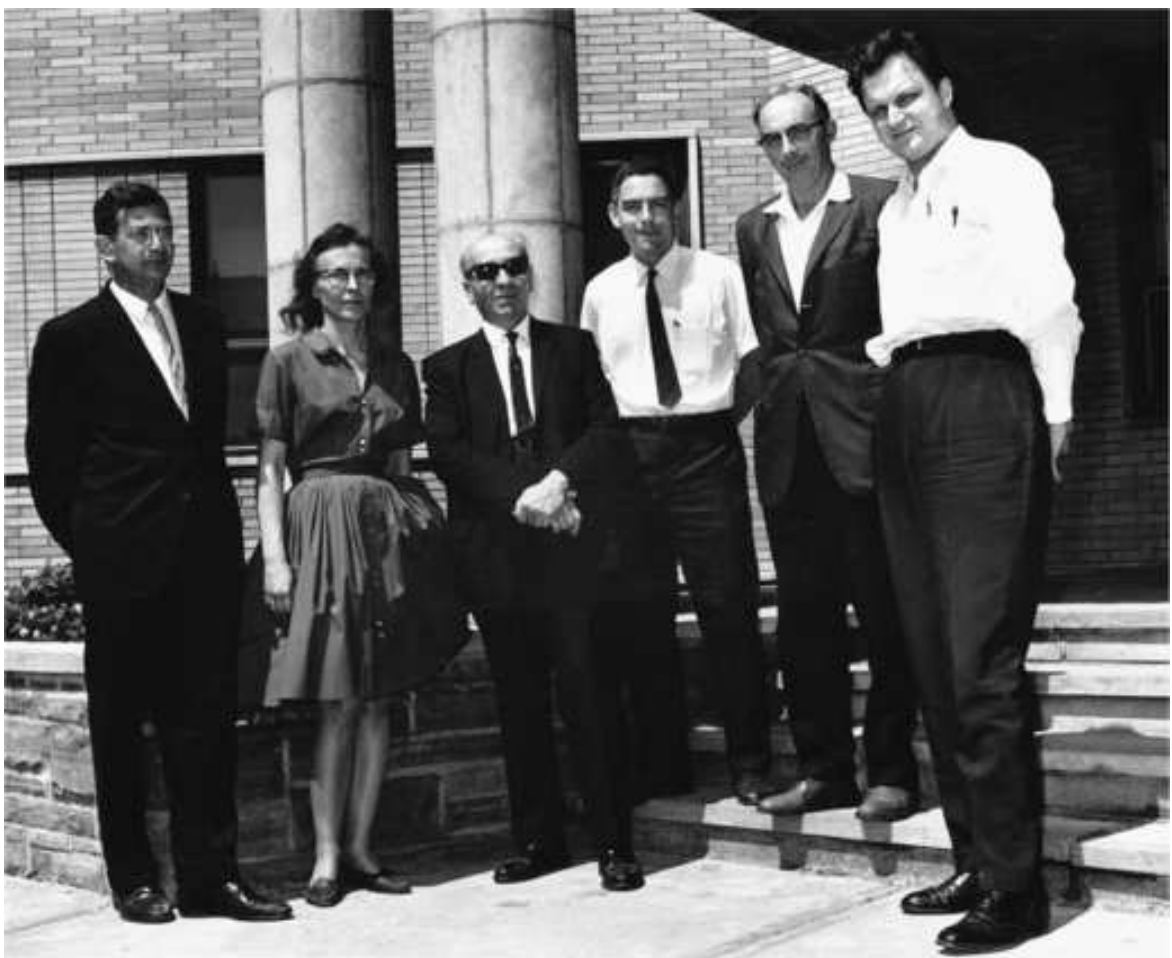

Fig. 3. *Montréal 1968. Séminaire de Mathématiques Supérieures. The six speakers: Sam Karlin, Constance van Eeden, Mark Kac, PJH, Lucien Le Cam, Jacques Neveu.*

**AB:** You mentioned that Beno Eckmann was not at all in favor of you going into statistics.

**PH:** No, he was not. It was Stiefel and Saxer who tried to persuade me.

**AB:** And yet Beno Eckmann somehow recommended that you go to Battelle to see some other type of problems...

**PH:** Battelle, oh, that is a different story. Eckmann had managed to get a contract from the U.S. Army Research Office, but he couldn't be the principal investigator as professor of ETH; so he had to have a pro forma principal investigator, and that was a student of his, Heinrich Kleisli. But Kleisli got an offer from a university in Canada, and Eckmann urgently needed a new principal investigator of that project, so he put me in charge.

**AB:** What kind of experience was Battelle?

**PH:** The project was really in a theoretical physics group, a very interesting place, with interesting people, but my work was on homological algebra. Battelle was an umbrella for many different projects. The U.S. Army Research project was channeled through Battelle. This was a short stint, though, less than a year, from October 1960 to the following July.

**AB:** So then Eckmann's design with you would have been what?

**PH:** Eckmann thought that I should stay in topology, and then I would probably have gone to Berkeley also, but a floor lower in Campbell Hall. Eckmann thought I should first publish some more papers in topology before I branched out or switched fields.

**AB:** Few statisticians these days study category theory or topology at this level, but can you give us a rough idea what sorts of things you worked on?

**PH:** Originally Eckmann had suggested a problem to me that was fairly concrete on the interface between homotopy theory and homological algebra, and I worked very hard on it for about a year. Then I decided that the problem was ill posed, that it didn't have a solution; I never convinced Eckmann, but I convinced myself. Eckmann at the time was actually interested in another problem, a curious analogy between topology and algebra. There were certain similarities between the two fields that Eckmann described with the term "duality." I realized if one rephrased the matter in terms of categories, things became identical. It only depended on how one interpreted the objects, the morphisms as continuous maps on the one hand and as algebraic maps on the other. This fitted in with Eckmann's views, and it became my thesis.

**HK:** In addition to van der Waerden and Eckmann, you mentioned some other mathematicians who were influential in your life. Could you comment on Stiefel?

**PH:** Stiefel originally made himself a name as a topologist. Stiefel manifolds are named after him. He had been a Ph.D. student of Heinz Hopf. (Incidentally, Hopf was the second reader of my thesis.) When Stiefel was a young faculty member at ETH in the 1940s, somebody had to teach applied mathematics, and the lot fell to him. It seems he liked it. Besides, he became the co-inventor of the conjugate gradient method. He was also the driving force behind computing at ETH in the early days. I will say more about him when we talk about computers.

## BERKELEY, CORNELL AND THE 1964 PAPER

**HK:** So you came to Berkeley, after reading van der Waerden's, Wald's and Doob's books. How did you decide on a topic to work on?

**PH:** I have no recollections of what I was planning to do. Peter Nüesch, who was then a graduate student at Berkeley, picked us (my wife Effi, myself and our little son Thomi) up from San Francisco Airport (it was in 1961), and some years ago he claimed that, while we were driving to Berkeley, I had told him that I wanted to work on a theory of robustness. I can't confirm this, but it could be true. Van der Waerden was interested in nonparametrics, and he was worried about the reliability of distributional assumptions, and so was I. Van der Waerden's book and my study of Wald's decision theory combined to convince me that, if it were possible to build a theory of robustness, it would be through decision theory. These thoughts did not gel till about a year later when I found that certain M-estimates have an asymptotic minimax property—which became the nucleus of my 1964 paper.

**HK:** Were you influenced by Tukey's 1960 paper "Sampling from contaminated distributions"?

**PH:** Certainly. I don't know when I read it first; it must have been pretty early at Berkeley. I think I didn't meet Tukey until after I had finished writing the 1964 paper. The paper was submitted sometime in spring or summer 1963. That summer we moved to Cornell. I think that's when I met him, sometime in 1963/64, on the East Coast.

**HK:** Some people say that the idea of robustness, that you pay a price up front as insurance against things going really bad, is a very Swiss kind of mentality.

**PH:** It was Frank Anscombe's idea, though. It might have been in his 1960 paper on the rejection of outliers, or it could have been already in an earlier of his papers; as far as I remember the insurance idea is Anscombe's. So there isn't much Swiss about it really...

**HK:** With whom did you collaborate in Berkeley? Who was influential in writing your 1964 paper?

**PH:** Erich Lehmann. I believe Lehmann was editor of the *Annals* but I am not sure anymore; he may have been editor earlier.[2] Anyway, Lehmann was the natural person to get advice on how to publish a paper in the *Annals*.

**HK:** Were there any difficulties, such as referees missing your key points?

**PH:** Not in my recollection. Lehmann gave some very good advice, telling me that I should submit a paper in a preliminary version that was too long. This has two advantages: First, referees would understand what it was about and, second, they would recommend shortening it, so I could start revising it right away after submitting it.

**AB:** Do you remember anything from the referees' comments?

**PH:** No, I don't.

**AB:** So it must have been smooth sailing.

**PH:** I guess so. I have no idea who the referees were either.

**AB:** More about Berkeley: what kind of interactions did you have with Lucien Le Cam?

**PH:** I was sitting in Le Cam's course on decision theory where he did comparison of experiments. You [to AB] got into something like this later, didn't you?

**AB:** Exactly. He introduced $\epsilon$-sufficiency also in a 1964 paper, and you suggested that there should exist a link to robustness, which was indeed the case.

**PH:** Le Cam's course was not easy, so I reworked the material. Subsequently I taught at Cornell for a year and gave a graduate course on decision theory.

It so happened that Larry Brown was one of the students, and he obviously stood out.

**AB:** One last question about Berkeley: was Jerzy Neyman there?

**PH:** Neyman was still active; I think he officially retired while I was there, but he continued to be around all the time.

**AB:** How did you experience him?

**PH:** I had little contact with him, but his hold on the place was immediately apparent and very amusing. Neyman still *was* the department. At the beginning of my stay he was on leave and the department was sleepy, but when he returned the atmosphere came alive from one week to the next. The place burst with activity and one could see, for example, graduate students busily addressing envelopes for fund raisers for Martin Luther King. It was indeed amusing.

**AB:** So he brought energy to the place...

**PH:** He was the driving force behind the department. I have never had a similar experience again, an eruption of a place from sleepiness to bursting activity.

Other people were less overtly visible because they had odd working hours, such as Joe Hodges who worked during the night, and Lehmann you best caught after class. Le Cam was very nice, but you had to approach him. The only professor who was regularly in the coffee room was Michel Loève, and with him you could discuss anything. David Blackwell came sometimes. The coffee room also housed a reprint collection from which I learned much about Tukey.

**HK:** Did you have contact with graduate students and other postdocs at Berkeley?

**PH:** Let me jog my memory. I shared the office with Don Burkholder who was on leave, and among the graduate students were Peter Bickel and Richard Bucy of the Kalman–Bucy filter. I believe the Kalman–Bucy filter had been invented the year before. While I was there David Freedman, coming from Princeton, was made assistant professor.

**AB:** How did you get to Cornell after Berkeley?

**PH:** I could have stayed at Berkeley because I had first a one-year Swiss National Science Foundation fellowship and then a two-year fellowship from the Miller foundation. I wanted to experience one other place as well, and I wanted to get some teaching experience, partly out of curiosity and partly for career reasons. Jack Kiefer offered me a visiting position at Cornell.

---

[2] Actually, Lehmann had been editor 1953–1955; from 1961–1964, the editor was Joseph L. Hodges, also at Berkeley.

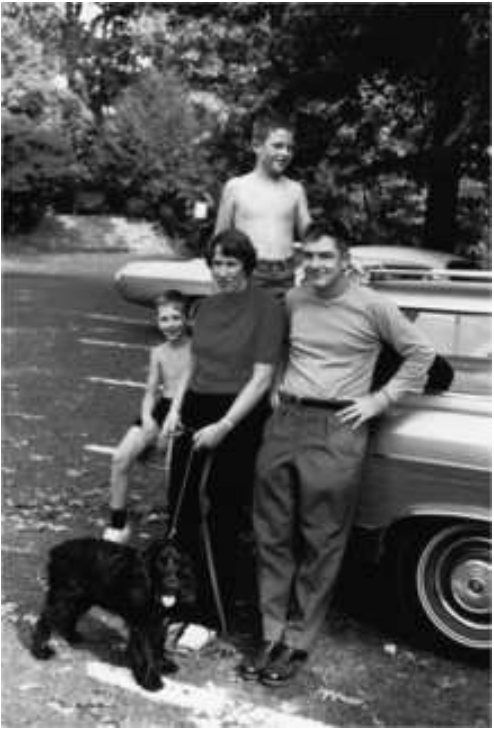

Fig. 4. *Princeton, Fall 1970. With sons Thomas and Niklaus, and dog Tappi.*

**AB:** How did you know Kiefer?

**PH:** I had met him in Berkeley.

**AB:** And after Cornell you got a call from ETH?

**PH:** Yes, this was in spring 1964. I was getting nervous over my J-visa that was to expire after three years. It couldn't be extended, and the rules mandated two years in the country of origin. So the call from ETH came in handy.

## ROBUSTNESS AFTER THE 1964 PAPER

**HK:** Can you describe how the research continued? You had this 1964 paper coming out, and then I think the next step was the robust testing approach?

**PH:** Yes, the asymptotic aspect of my robustness theory was unsatisfactory because it makes a major difference whether one has, say, 1% contamination and a sample size of 10 or of 1000. So I was very happy when I managed to get exact finite sample results. I think among those papers I like the one in the "Zeitschrift" best.

**AB:** ...the one on "Robust Confidence Limits." Earlier you had also published a robust version of the Neyman–Pearson lemma.

**PH:** Yes, and that paper led on the one hand to the paper in the "Zeitschrift" and on the other hand to the paper with Volker Strassen.

**HK:** How did you meet Strassen?

**PH:** I may have met him at the 1965 Berkeley Symposium, and I tried to get him to come to Zurich. ETH, however, was problematic because he would have to teach calculus for scientists, which he refused to do. ETH was not willing or not able to give him another appointment. I then lobbied van der Waerden to get Strassen to the University.

**AB:** ...and that worked?

**PH:** It worked, but both Strassen and I made it very clear that he would probably not stay in probability, even though they would hire him as a probabilist. Fortunately, the University didn't quite believe him, but it came true anyway. Part of Strassen's charter would have been to do statistical consulting, but he refused that, too, so he negotiated that he could bring along a young person to do the consulting, namely, Frank Hampel. Both Strassen and Hampel must have arrived in 1968. We all were in contact with each other, had common seminars on computational complexity and on robustness, and it made life in Zurich a lot more interesting.

**AB:** When you talked to Strassen, did you discover that there was a possibility of doing something with Choquet capacities to generalize your robust neighborhood tests?

**PH:** I had read Choquet's paper in connection with Markov processes and potential theory, but this was really Strassen's idea.

**AB:** So you told him about what you did and...

**PH:** If I remember correctly, he had read my paper on robust tests. He drew a connection to his thesis (where he had used capacities to formalize inaccurate knowledge of probability measures on finite sets), and he told me that this was essentially a capacity argument. We had many discussions— I don't remember when these discussions started, some time before he came to Zurich—but since we did not manage to extend the theory beyond finite sets, we let the work sit for a while, and in the end, in 1970, Strassen said we should make a final effort to generalize it and write it up. I remember that we talked about this on a long walk in the woods above Zurich, a few weeks before I left for Princeton. I made the major push when I was at Princeton for the robustness study. I sent a draft to Strassen, but by that time he had fully moved into computational complexity and algebraic geometry, and he was no longer interested. The draft was sitting, and in the end we both had forgotten the details. I am still a bit unhappy because some errors made it into the publication.

**AB:** Well, that became grist for others. At any rate, this is another side of robustness that is much less known than the 1964 paper with its asymptotic theory: robust tests where both $H_0$ and $H_1$ are contamination neighborhoods or, more generally, Choquet capacities, and based on these tests you invented "Robust Confidence Limits" (1968). This looks like a prettier and more satisfying theory.

**PH:** Indeed. Its importance is the following: it shows that essentially the same procedures that are asymptotically optimal for symmetrically contaminated distributions are also optimal in a *finite sample sense* and for *arbitrary asymmetric contaminations*.

**AB:** The symmetry assumption of the asymptotic theory was indeed a longstanding criticism.

**PH:** Yes, and I think it is still not widely known that the finite sample optimality results about robust confidence limits essentially finished up that question. Maybe I should have pressed this side of robustness a little more. I was never sure whether the theory of my 1964 paper was the real thing, but it did give the idea of robustness an element of respectability, which is all that I could have hoped for. I feel more strongly, though, that the paper on robust confidence limits is the real thing.

**AB:** But in addition to theory you also came up with complete algorithms.

**PH:** Yes, but people always thought that I was a bloody theoretician. Yet, there is merit to theory in that optimality results give us some sign posts that here, in this direction, one cannot go further. Optimality tells us under what conditions there are limits to how well a method can perform, and if a method is close to but not perfectly optimal, this might be as well.

Back to robustness theory: There was criticism of the robust minimax results also because of the form of the least favorable distributions that were pieced together and hence not analytic. I found this an unworthy objection because, if the least favorable distributions are realistic, they make sense, never mind their analytic form. In fact, later I was pleased to see that least favorable distributions were often closer to observed data distributions than the normal distribution.

Which brings me to real data... I had always been interested in data analysis, and I got interested even more during the robustness year at Princeton in 1970/71. That, I think, was the turning point: I essentially got out of robustness and moved into data analysis. I still had to write a book on robustness...

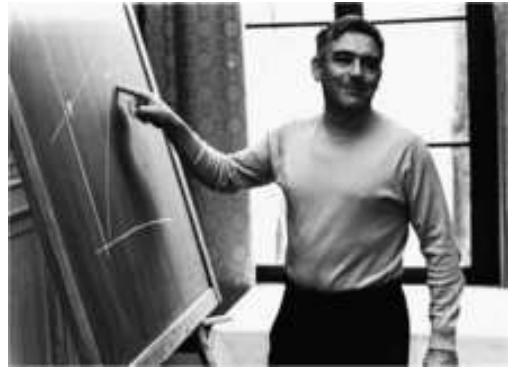

FIG. 5. *Schloss Reinhardsbrunn, German Democratic Republic, 1975. Peter discusses leverage points.*

**AB:** ...and you still did other work on robust covariances and robust regression, the Wald lectures, and so on.

**PH:** Sure, I was still in it, but I really had my mind set on data analysis. Sometime in the 1970s I realized that I was intellectually going in circles. That is, I would have an idea, and somehow I would fall into the same track, like a broken record: every problem turned into some kind of minimax problem. As the saying goes, if you have a hammer, every problem looks like a nail.

## THE DEVELOPMENTS OF COMPUTING

**AB:** Can you tell us more about how you first got involved with computers?

**PB:** I got into computing I think in 1956 when ETH had acquired the new ERMETH computer built from vacuum tubes, and Rutishauser gave a course on programming it. Computing at ETH started with Stiefel. At the end of the war, he had heard about the "Zuse computer" that was stored in a shed somewhere in southern Germany, and he managed to get hold of it and move it to Zurich. Stiefel then formed a small subgroup for computing research within the applied mathematics group.

**AB:** Tell us more about the Zuse computer.

**PH:** It was a very amusing piece of hardware. It was partly made from the tin of war-time quality tin cans and therefore had many problems. It was not mechanical but based on relays. When Zuse's machine was demonstrated to us I was intrigued but not impressed. It took about 10 minutes to solve linear equations with three or four unknowns, and I felt I could do that faster by hand. This must have been toward the end of my first term at ETH, in early 1955.

A few years before,[3] Stiefel had sent two of his younger people to the United States to learn about computers: an electrical engineer, Ambros Speiser, and a mathematician, Heinz Rutishauser. When they came back they designed a computer made from tubes and diodes. It became operational in 1956. It was called "ERMETH," and it was one of the first floating point machines. It was not binary but decimal with double precision floating point numbers. It had a drum memory as its only memory, with 10,000 words, in all about 50 KB of memory.

**AB:** ...which must have been big at the time.

**PH:** That was big, indeed. The computer was pretty slow, though. An addition took 4 msec or so. There was no assembler, so one had to program in actual machine code. I played with ERMETH for fun, but then my wife began to use it in a bigger way.

**AB:** Your wife, Effi: you met when...?

**PH:** We met in high school and we married shortly after our ETH diplomas. She was in crystallography, and crystallographers have been big computer users from the start.

**AB:** So you were both early computer users.

**PH:** Yes, we were indeed both early computer users, yes. But Effi was really the big user. At times she had jobs that lasted 24 hours. This meant that she had to attend to the machine and restart it every few hours because it would break down frequently. One had to program very carefully so one *could* restart without going back to square one. I got into computing and data analysis through Effi.

**AB:** What type of data analysis would she do?

**PH:** Three-dimensional Fourier syntheses and least squares problems, nonlinear weighted least squares. In her thesis she dealt with maybe 37 unknowns and 1000 to 2000 observations. For her final results she considered using ERMETH but saw that it was too risky: the machine was too slow, too unreliable, and restarting was difficult. So she ended up doing her least squares problems on a computer at CERN in Geneva.

**HK:** Could you also describe for us the developments in computer software and programming languages, how you experienced them?

**PH:** Programming on the ERMETH was fun. There were three levels of languages on top of each other: First, one wrote a program in a flow chart language, then one translated that flow chart by hand into a kind of assembly code with symbolic addresses, and then one translated that assembly code into machine code.

**AB:** ...manual, not automatic?

**PH:** By hand. Subsequently, the problem was making changes in the program because usually the changes were not local and the addresses were absolute. Based on his early experiences with computing and numerical analysis, it was natural that Rutishauser became one of the driving forces behind the development of ALGOL 60, which at the time was still called ALGOL 57. One should know that Rutishauser had written the first paper ever on compilers in 1951, his *Habilitations Schrift*, published in 1952. The point of his paper was to propose a language that made it possible to describe numerical algorithms unambiguously. He was actually not interested in compiling per se, but machine compilation would show that the description was complete and consistent. In 1960 ALGOL 60 came into being. As Rutishauser told the story, there was a large committee that had agreed on the language, but in the end Naur completely rewrote the final document, which became then known as the ALGOL 60 report. Various participants of the ALGOL conference, attended by maybe a dozen people, agreed to write compilers or have compilers written. In Zurich it was Hans Rudolf Schwarz who wrote an ALGOL compiler. I tried it out but ended up unimpressed again: a little program for a Jacobi eigenvalue problem that I had written took almost an hour to compile.

The program is of some interest because it was part of a data analysis problem of Effi's. The problem was to map intensities measured from different photographs to the same scale, as when one has inexact crystallographic data. This kind of problem is nowadays called a Procrustes problem. To me this was an opportunity to try out the ALGOL 60 compiler. I was probably the first ALGOL 60 user outside the applied mathematics group of Stiefel.

The next step in the development of languages that I experienced was Fortran. In late 1960 or early 1961 Effi and I attended a Fortran programming course at CERN in Geneva, programming on an IBM 704. The computing environment was still primitive. For example, if one wanted to write something to tape, one had to get the tape running before one began to write. So one had to estimate how much earlier one had to start it. But more importantly, Fortran compilers were faster.

---

[3] 1948/49.

The next step was several years later, it must have been 1967, when Niklaus Wirth came to Zurich. It so happened that our neighbors were on an extended absence to the United States; Wirth ended up renting their house and we became neighbors. At the time Wirth was developing the Pascal language. I looked at Pascal but was disappointed because in strict Pascal array bounds were fixed, which is why I didn't bother to learn Pascal. Neither could I convince Wirth that he should provide greater flexibility.

At that time we had a very efficient ALGOL compiler on the CDC 1604 computer. Unfortunately, ALGOL 60 then got killed by its successor, ALGOL 68, which was a big fiasco because it was so complicated that it never took off. It had what Donoho used to call the "second-generation syndrome": start with a good product, then follow it up with an over-designed second-generation product, only to see it killed. This happens often, and it did happen to ALGOL 68.

Next was the appearance of the C language, but I never took to it because it was too error-prone. So I stuck to Fortran, which was sufficient for my type of computing. Later I had occasion to program in C, but, matter of fact, I have used more often f2c, the Fortran-to-C cross-compiler, which seems to work better than most Fortran compilers.

**AB:** So you have seen computing from the beginning…

**PH:** Speaking of beginnings, in the 1970s we began to experiment with interactive graphical data analysis. This changed my views of computing rather dramatically. I soon realized that we needed a data handler, that is, a language in which we could not only write programs, but whose command lines were capable of immediate execution. BASIC and APL were such languages. I liked BASIC for writing little things, especially string manipulations, but we needed an array-oriented language. As for APL, it was a "write-only" language, I could write it, but I could not read it, and I may be a kind of expert in such matters. I never understood Anscombe why he had written his programming examples in APL. When I asked him he confessed that he couldn't read his own programs either after half a year!

I guess that I should expand on our experiences with data handling. After I had moved to Harvard in 1978 and started a graphics project there, Donoho proposed to revive the ISP language as a data handler for the project. He had co-developed ISP at Princeton. Donoho left in 1983, and then Effi and I extended it, and to this day I am using ISP for my own purposes. It is still easiest for me to use it both interactively and to do ad hoc programming. I never got used to S. Neither to Matlab, although I am familiar only with early versions. At any rate, I am still using ISP which is great for me to improvise solutions to nonstandard problems.

**To AB:** How do you deal with nonstandard computations?

**AB:** I grew up with S and later grew into R, so I know all kinds of tricks. Maybe it depends on how one grows up. Maybe it is a subjective thing, determined by life history.

**PH:** Your mother tongue is what you stay with.

**AB:** Yes, but one has to have one mother tongue that one knows in and out, and so students I think have to have one.

**PH:** Another question is, what should you learn as a student? In my experience, one area that is generally underemphasized is reformatting of data. Often one has to know tricks to do reformatting, and usually one has to program these things in a low-level language, in particular when facing binary formats that require not only manipulating bytes but bits. We ended up putting a bit handling facility into ISP. Most people probably write C code to do bit handling. We thought a long time about how to do it, and now I know I can do essentially everything with our bit handling facilities.

**AB:** You made it a high-level problem…

**PH:** Bit manipulations are particularly useful when apparent data corruptions turn out to be something else. Once we met a case where a programmer had put the main information into 7-bit ASCII but then squeezed additional information into the eighth bit. We experienced something of this kind also in the children's growth data at the University of Zurich, based on a 20-year longitudinal study on which Werner Stuetzle did his thesis. The background was as follows: Certain events in bone development were supposed to occur between ages 5 and 9, and so a single decimal digit position had been reserved for age. Then a few children had this development at age 10 or 11. The punch-card operators made a very intelligent decision: they encoded the information allowing letters in addition to digits. This was not documented and we didn't know when we started to analyze the data. The letter codes were thrown out by a data-checking mechanism as punching errors. Then the person in charge of transferring the punch-card data to tape (it may have been Theo Gasser) actually went back to the original handwritten data sheets, which still existed, and realized what was going on. And we managed to read the cards properly.

Another, published, example is in a famous JASA paper by Coale and Stephan,[4] a very worthwhile piece. The authors looked at 1950 census data and realized that there were quite a few 14-, 15-, 16-year-old widowers. Coale and Stephan, in something that reads like a detective story, describe their discovery that a few thousand punch-cards must have been punched with one column shift, so a 32-year-old head of household was turned into a 13-year-old widower, and a 42-year-old into a 14-year-old widower, and so on. Apparently, 13-year-old widowers were automatically thrown out as errors because it was legally impossible to marry at age 13, but one could marry at age 14 in some states.

**AB:** This actually brings up the more general problem that analysts of large datasets can easily be misled by artifacts and data corruptions that are difficult to see. Datasets can be strange for systematic reasons such as you just mentioned. So is there any wisdom other than "Well, be careful"?

**PH:** I think the specific lesson you can draw from our children's growth data and from Coale and Stephan's widows and Indians is that if there are clusters of bad data, one has to dig in because very often bad data have meaning.

## DATA ANALYSIS AND DATA VISUALIZATION

**HK:** Can you tell us something about your views and experiences with data analysis?

**PH:** The problem with data analysis is of course that it is a performing art. It is not something you easily write a paper on; rather, it is something you do. And so it is difficult to publish.

**HK:** If you analyze data in a subject area, don't you publish it there, not in mainstream statistics?

**AB:** Which you did, too. You worked on EEGs with Gasser, pretty much after the Princeton study; actually Gasser was with you in Princeton.

**PH:** Yes, but the EEG work started earlier, with the Cooley–Tukey algorithm, which must have been in 1965. The University of Zurich had an EEG project in collaboration with ETH mathematicians. They, however, were uncomfortable because there was too much statistics involved. They suggested that I take over. The project required stationary time series analysis. At the time I had just learned from a talk at an ISI meeting in Belgrade that some such thing as a fast Fourier transformation had been invented, and that it worked best with powers of two. This was enough of a hint to allow me to program it up myself. The regular Fourier transform with its n-square computational complexity was much too slow to compute on a CDC 1604 computer [CDC = "Control Data Corporation"]; it required hours and hours. This reminds me of an amusing anecdote: The computer operators had a microphone hooked up to the top bit in the accumulator so one could hear what was going on in the computer. Unfortunately, the fast Fourier transform had the property that it would make a howling "oooh oooh oooh" sound, which was usually an indication that the currently running program was stuck in an infinite loop. As a result the operators often terminated the program prematurely, which didn't speed things up either.

**AB:** Look at that! An early example of sonification...

**HK:** Later you also looked at higher-order spectra of the EEGs. Was this successful?

**PH:** The director of the project, Guido Dumermuth, was very much interested in higher-order spectra, so I looked into it. We didn't get very far, though. We calculated some, but they were not easy to interpret and rather sensitive to artifacts.

**AB:** That sounds like another robustness problem.

**PH:** Possibly; one had to be very, very careful about tapering and the like to avoid artifacts.

**AB:** So then in the seventies you turned yourself loose on computing.

**PH:** Effi had a large role in that. When we were at Princeton in 1970/71 she worked in Langridge's molecular biology lab. Langridge had received new equipment that was very fancy for the time: a DEC-10 [DEC = "Digital Equipment Corporation"] and an Evans & Sutherland vector graphics display, which had just arrived when we came—the naked display hardware, without any software. Effi got a job in this lab as a postdoc. Langridge told her something along the lines, "Here is this great new equipment, can you do something with it? And by the way, we have crystallographic data on transfer-RNA, and also a geometric model of transfer-RNA." Effi was curious enough to test whether the model would fit with the crystallographic data, and that meant among other things fitting a molecule into its observed crystallographic dimensions. In essence we wrote a program for molecular packing. In retrospect I'd say it was a fledgling expert system.

---

[4]COALE, A. J. and STEPHAN, F. F. (1962). The case of the Indians and the teenage widows. *J. Amer. Statist. Assoc.* **57** 338–347.

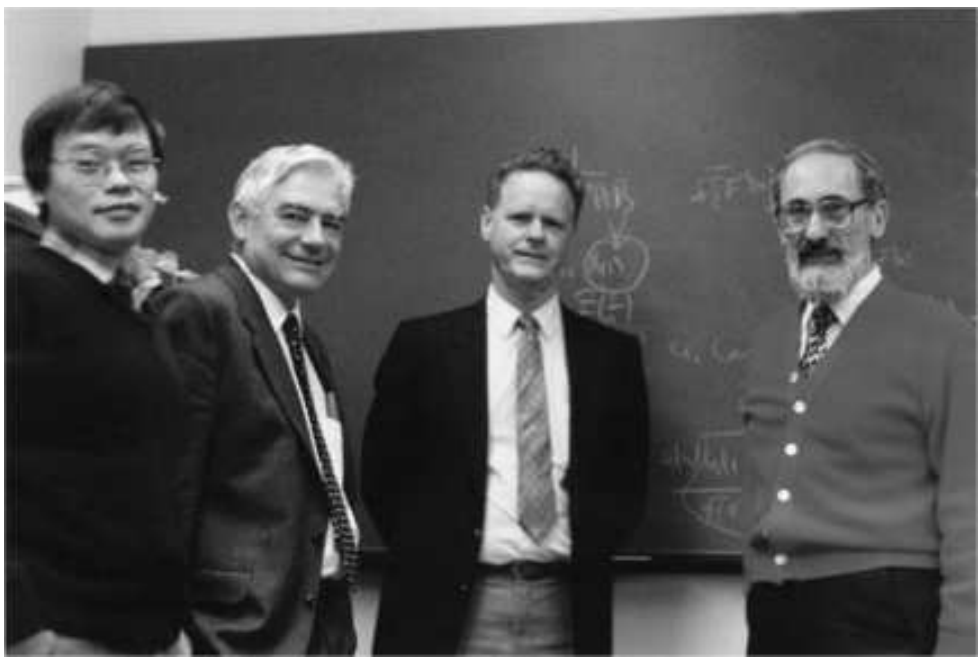

Fig. 6. *Harvard 1986, on the occasion of Fred Mosteller's 70th birthday. Shaw-Hwa Lo, PJH, Art Dempster, Herman Chernoff.*

**AB:** Meaning you used heuristic rules?

**PH:** One had to input some initial coordinates. Then the program would try to improve the packing of the molecules, following some simple rules which we bettered over time. When it became stuck, it got back to you, and one had to figure what might have gone wrong. For example, some protrusion of one molecule might have gotten stuck between protrusions from other molecules, in which case one had to view the situation in 3-d, pull them apart and restart packing. I found the problem very interesting.

**HK:** So then you thought this equipment would also be useful for statistics?

**PH:** Indeed, and for data analysis in general. I agree with Tukey in that I take data analysis to be the larger thing, comprising statistics. I realized that with high-powered graphics equipment one could do things one couldn't do otherwise. Until then I had looked at graphics as just a toy.

**AB:** Did you learn about the PRIM-9 movie at the time?

**PH:** The PRIM-9 movie, oh, it was made later. We were beyond PRIM-9 because of Langridge's equipment. Tukey sometimes visited and watched when we were there—he probably got the idea for PRIM-9 from there. Later he would visit Jerry Friedman at SLAC (the Stanford Linear Accelerator Center) where he had Mary-Ann Fisherkeller program what looked to us like inadequate equipment. The things one could do with Langridge's equipment were much more advanced.

**AB:** Tukey could have stayed right there in Princeton?

**PH:** Actually, he couldn't; Langridge wouldn't let anybody outside of his group use the equipment.

**AB:** Oh, so Effi was lucky to be part of the group.

**PH:** Yes, she had an appointment there. Also, the few people who knew how to operate it were using it almost full time. Effi went there mostly during the night.

**AB:** Tukey wasn't the operator who would try to get his own equipment, so he went to SLAC instead?

**PH:** The equipment was also expensive. You may not remember the prices we paid for the interactive equipment we purchased for ETH.

**AB:** ...hundreds of thousands...

**PH:** I seem to remember the budget was about 3 million Swiss francs...

**AB:** ...for a DEC-10, a PDP-11 and an Evans & Sutherland display, which is the equipment that you later purchased at ETH, after much difficulty.

**PH:** Yes, this brings up the point that the equipment at Princeton was much easier to operate because there was no PDP-11 mini-computer between the DEC-10 and the display. For obscure reasons—I think it was price—Evans and Sutherland decided to hook their display to a PDP-11, which slowed down the graphics and complicated the programming.

**AB:** Ironically, soon after the arrival of the equipment at ETH, you left ETH for Harvard. Werner Stuetzle programmed PRIM-ETH on the equipment in Zurich, and you started work on PRIM-H.

**PH:** I should explain the irony. Our proposal had prevented ETH from missing out on an important development in computing. The irony was that instead of appreciating it, the administration was venomously furious that we had dared to interfere with entrenched powers and to compete for funds with the batch computing establishment. Just then, Harvard made me an offer, and I was glad for it. At Harvard I got access to practically the same equipment, in the chemistry department, after relatively smooth negotiations. The graphics software was mostly done by Mathis Thoma and the data handler software by David Donoho.

**HK:** At that time you also got interested in projection pursuit.

**PH:** By then, I had realized that it was difficult to search visually for interesting views of more than three-dimensional data. And again, there is this problem that data analysis is a performing art, and in order to write papers one has to get into theory. One possibility that came along was projection pursuit.

**AB:** You worked on different projection pursuit criteria?

**PH:** Yes, with Donoho. It was very interesting, and he had some brilliant ideas in that area. One problem was to describe the invariance properties projection pursuit indices should have and then figure out which indices had them.

**AB:** Donoho didn't publish on projection pursuit himself.

**PH:** I don't think that he did. He would move on...

**HK:** You said that data analysis is a performing art. Can you describe to us some projects or data that you analyzed that were particularly interesting and what areas they came from?

**PH:** I like analysis of out-of-the-ordinary data. An example, currently of concern to me, are astronomical data where one faces peculiarities such as the length of day and the irregular rotation of the earth. Such data require different time scales, one being the uniform, dynamical time scale that underlies the gravitational theory of the solar system, and the other being the civil time scale that relies on the rotation of the earth. Because the rotation of the earth is irregular, the length of day changes systematically over time as well as randomly. If one tries to extrapolate time for the calculation of ancient eclipses, one has to have an idea how big the extrapolation error might be. Extrapolation always involves a model, so one has to build a model for the rotation of the earth, check it for adequacy, and estimate its parameters. I rather like such intricacies. This was, on the one hand, an analysis of the data and, on the other hand, an effort in model building. It turned out that a Brownian motion model fitted the data pretty well. I wrote a little paper which I buried in a Festschrift, and lately I have been working on an update which I would like to publish sometime.[5]

As for data analysis in general, by 1990 I felt that I knew enough about the opportunities and technicalities of interactive data analysis and graphics, and I drifted into what might be called the philosophy of data analysis. The two papers of mine that you published as editor of the *Journal of Computational and Graphical Statistics* (JCGS), I like them quite a bit,[6] as well as a third one on strategy in data analysis.[7] I once wanted to remake all three of them into a kind of a prolegomenon to a book on data analysis.

**AB:** You also wrote some comments on the past, present and future of statistics.[8] Would you say anything differently from what you said then?

---

[5]Now published as: Modeling the length of day and extrapolating the rotation of the Earth (2006). *J. Geodesy* **80** 283–303.

[6] Massive data sets workshop: Four years after (1999). *J. Comput. Graph. Statist.* **8** 635–652. Languages for statistics and data analysis (2000). *J. Comput. Graph. Statist.* **9** 600–620.

[7]Strategy issues in data analysis (1997). In *Proc. of the Conference on Statistical Science Honoring the Bicentennial of Stefano Franscini's Birth* (C. Malaguerra, S. Morgenthaler and E. Ronchetti, eds.) 221–238. Birkhäuser, Basel.

[8]Speculations on the Path of Statistics (1997). In: *The Practice of Data Analysis, Essays in Honor of John W. Tukey* (D. R. Brillinger, L. T. Fernholz and S. Morgenthaler, eds.) 175–191. Princeton Univ. Press.

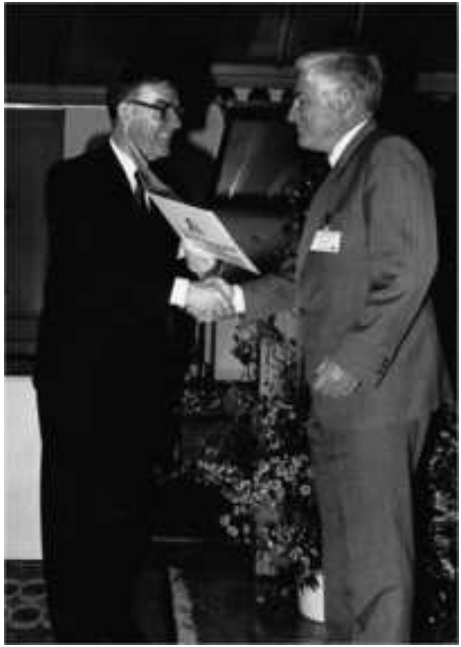

FIG. 7. *Tegernsee, 1990. Peter receives Humboldt Award from Professor Reimar Lüst, president of Humboldt Foundation.*

**PH:** I don't think so.

**HK:** You mentioned this paper by Coale and Stephan (1962) that you find interesting. What other papers, if you look back, did you find particularly interesting?

**PH:** I should mention Tukey's "Future of data analysis"; the first few pages and the last few pages in my opinion are a must for every statistician. The part in between has lost interest with time, but the first and last few pages are here to stay. The most remarkable aspect is that this paper was published by the *Annals*.

**AB:** Earlier we talked about the question how difficult it is to see problems in large datasets. Is there anything else to say about massive datasets?

**PH:** Yes, there are more aspects. A problem is that the statistician of old times who analyzed data by hand would notice if something was amiss, whereas the modern statistician sees masses of data filtered through computer manipulations that may conceal data problems. Up to a few megabytes, computer graphics can reveal if something is amiss, but going beyond, it gets difficult to notice if something is even grossly wrong. This is, by the way, one of the problems with data mining. All too often when mining data one hits on trivial "nuggets." I recall the case of a data analysis problem that was part of the Ph.D. exam at Harvard. It was a discriminant problem where one had to discriminate between carriers and noncarriers of a certain genetic disease. A student found that the variable that discriminated best was "age." What he had found was that carriers and controls were not properly matched in age. Now, if one blindly runs a black-box algorithm on this problem, such as a genetic algorithm, the important, but trivial and possibly misleading, role of age will never be detected.

**AB:** This is a classical problem with a confounding factor. This is not something for which one can find an automatic solution.

**HK:** I imagine it must have been difficult for the student to prepare for such an exam.

**PH:** Yes, but it is very interesting to see what different people do facing the same data analysis problem.

**AB:** One of the difficulties for many students in statistics is vagueness, and this is what seems to have attracted you. Data analysis is partly an art, and that is very unsatisfactory for students who want to know the rules by which they can obtain good grades.

**PH:** Actually, I experienced the same in mathematics when I taught a calculus course at Cornell in 1963. Some students complained that in high school, mathematics had been very clear: there were clear problems with clear solutions. And now things were so vague and unsystematic, one could do things so many different ways.

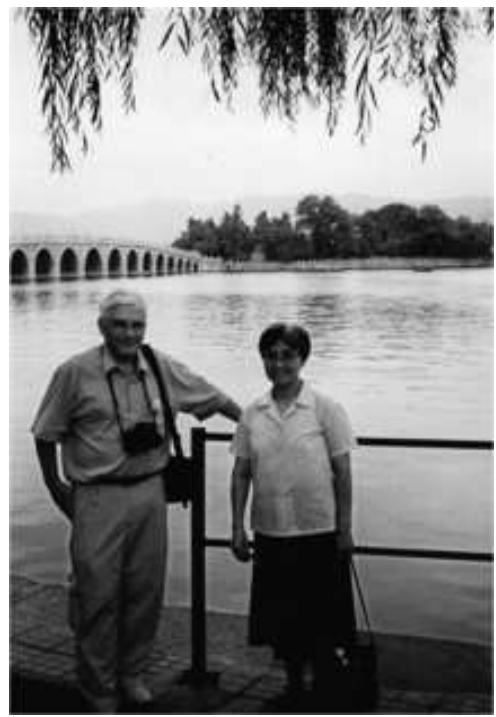

FIG. 8. *Beijing, 1997. Lecture tour in China. With the host, Professor Guoying Li, Academia Sinica.*

**AB:** ...and that was a complaint.

**PH:** As you said, students have problems with the art aspect.

**AB:** Are we digging in the wrong pool of students?

**PH:** Not really; as a student you have to get used to the art aspect, and we have to teach it. I just read a little paper by Steven Weinberg, the physicist, on the problems of teaching basic physics to nonphysics students. It seems to be very, very hard. I think they have a similar problem.

## BABYLONIAN ASTRONOMY AND ASSYRIOLOGY

**HK:** Maybe we can switch topics. We have seen from your list of publications that you have worked also very much on Babylonian astronomy and Assyriology. Maybe you could describe how you got interested in these areas and what you achieved there.

**PH:** Yes, I got into it in a funny way. In the "Gymnasium" (college-bound high school), when I was about 16, I read everything I could find on physics and astronomy and the like. I had gotten hold of a calculus book from which I learned calculus, then I got on to more advanced topics, and by the end of the year I had read something like Hermann Weyl's "Space-Time-Matter." Then I suddenly had it up to here. I knew I would be going into mathematics or physics later, but I just couldn't continue right now. I had to do something completely different. Somehow I ended up learning cuneiform...

**AB:** But you could have found some ancient Egyptian manuscripts, or you could have found literature on Brazilian tribes: instead you picked cuneiform...

**PH:** Oh, the reason for not choosing ancient Egyptian was that it didn't have vowels. I couldn't possibly learn a language if the vowels were not written and not known, so I switched to cuneiform, although in fact I did first try ancient Egyptian.

**AB:** I see you have here an Egyptian grammar.

**PH:** Yes, I tried later again to see whether I could, and I did, but at that time I couldn't, so it was cuneiform. I learned quite a bit of it when I was in the Gymnasium.

**AB:** It says here, in your CV, that you found the pertinent literature in an estate that was given to the library.

**PH:** The "Kantonsbibliothek" (state library) of Aargau had quite an extensive selection of books on cuneiform to start with. Toward the end of Gymnasium, I discovered that Neugebauer had published all the mathematical cuneiform texts available at the time, so naturally I got myself into that, too.

**AB:** Can you give us some background? There are different types of cuneiform, Sumerian, Assyrian, Babylonian,.... Is there sufficient commonality to learn them all?

**PH:** They have basically the same script, but their use is somewhat different, and they are different languages.

**AB:** How did you deal with the different languages? Did you pick a particular one?

**PH:** Akkadian I learned quite well, and then I learned some Hittite, and now I am dabbling in Sumerian. When I started at the University I found out that van der Waerden had written a book on ancient mathematics including Babylonian mathematics.

**AB:** ...without knowing the original texts?

**PH:** He had worked with Neugebauer's editions. Of course I approached van der Waerden, and I ended up writing a couple of papers on Babylonian mathematics and astronomy.

**AB:** Most undergraduates wouldn't even have a sense of what would be an interesting problem. How did you find interesting problems?

**PH:** John Tukey used to tell the fairy tale of The Three Princes of Serendip, who through sagacity would discover the most incredible things by the wayside. I guess that's the way to do it. Of course I heard that story from him only later. But Neugebauer's "Astronomical Cuneiform Texts" contained a couple of pieces that he hadn't really interpreted. So I had a careful look. I simply tried to go beyond what other people had done in this case.

**AB:** What type of mathematics did you find in cuneiform?

**PH:** Ironically, I ran into a problem treated in Babylonian math at a Gymnasium class reunion that I attended two weeks ago. A classmate, who does gardening somewhere in France, faced the problem of how to divide a lot in the shape of a trapeze into three equal areas. She did it by eye-balling and checking, but she felt there should be a mathematical method. Amusingly, this exact problem is solved in a Babylonian mathematical text. It leads to a quadratic equation.

**AB:** Did the Babylonians actually solve the quadratic equations?

**PH:** Most of the texts have problems without solutions. I guess they belonged to the curriculum of students. In this particular case the solution was described step by step. Sometimes one encounters oddities, though. For example, in one case there was a solution with an error in that a multiplication with a power of 60 was missed; but then this error was "corrected" by another error in the traditional sexagesimal (base-60) number system.

**HK:** So you continued publishing in this area all your life?

**PH:** Roughly every 10 years I got an attack of cuneiform. The first serious one, I think, was as a Ph.D. student when I was helping van der Waerden writing his book. I contributed some chapters on Babylonian astronomy. Later, when I was at Berkeley, I sat in courses on Akkadian, and similarly when I visited Yale. A bigger effort took place when I was asked to write a review of a book by J. D. Weir on the Venus tablets, that is, the old Babylonian Venus observations and their use for dating the Hammurabi dynasty. This involved doing some calculations as well as some extensive programming to carry out these calculations. Then, one day in 1973 when I was sitting in the library, van der Waerden came along and said he had been asked to participate in a panel discussion on Velikovsky and he was not keen on doing it and whether I would be interested… I think he had argued with Velikovsky before.

**AB:** You need to open a parenthesis here…

**HK:** …most readers will not know who he is.

**PH:** Professionally, Velikovsky was a psychiatrist. Because of an interest in Moses and the story of the Exodus, he began working on Egyptian history. But his evasive behavior, and his refusal to listen to arguments, in my opinion meant that he was a charlatan when it came to history. He had a theory that in historical times there were big catastrophes: Venus jumped forth from Jupiter as a kind of comet, erring around in the solar system for a while until it settled in its present orbit. All this was supposed to have occurred sometime in the second millennium BC. Suddenly, these Venus observations from the second millennium took on special relevance. So I said to myself if I had to write this review and do some calculations, I might as well put the results to use on the Velikovsky issue. This would also speed things up by imposing a time limit. The panel discussion was to take place the following year in February or March.[9] Being on that panel was a temporary culmination of this foray into ancient cuneiform writing. Because of the big shots on the panel—Velikovsky and Sagan—also my name made it for once into the New York Times, as a "Professor of ancient history"!

**AB:** What was exactly the result?

**PH:** Based on the Venus tablets I argued that Venus was in its present-day orbit back in the first part of the second millennium, and there was evidence for Venus also from much earlier times.

**AB:** …and what was Velikovsky's basis for this theory?

**PH:** His theory was a sham; he had lots of footnotes and citations, making it look very learned, but if one checked the footnotes one found that they were either obsolete or didn't quite say what he claimed.

**AB:** So why would he make this splash? Why was he not just thrown into obscurity from the outset?

**PH:** For various reasons: Macmillan Publishers Ltd was supposed to publish his book, but Macmillan was also one of the big textbook publishers. Several of their authors then threatened they would no longer publish with Macmillan. Macmillan withdrew, and Velikovsky made a big noise about censorship. He stylized himself as being maligned for his views by the establishment, which attracted many people, including scientists, who felt they were not sufficiently recognized. At this AAAS-sponsored panel discussion, the audience was split—half was pro and half against Velikovsky—and neither side would accept any arguments. Both sides were convinced they were right. Carl Sagan was on the panel, and he was just as much a prima donna as Velikovsky. Velikovsky would say the theories must be wrong because they do not agree with his evidence, and Sagan would say Velikovsky's data must be wrong because they do not agree with the theory. Rambling on and on, neither would keep to the time limits, but nobody dared stop them. Let's say, it was an interesting experience.

**HK:** But in the end your evidence solved the case, more or less.

**PH:** You do not solve such cases. Evidence is surprisingly irrelevant; opinions are stronger. What I claimed to be gross errors in a corrupt manuscript,

---

[9] It took place at the AAAS meeting in San Francisco, on February 25, 1974.

hard-core Velikovskians would take as evidence for the erratic behavior of Venus.

**AB:** But didn't someone pick up on your results?

**PH:** You mean people with a professional interest, Assyriologists and historians? Those people would be concerned about the precise date of the observations. Venus phenomena are fairly periodic; they repeat themselves almost exactly after 8 years except that there is a shift in the lunar calendar by 4 days. After 7 or 8 such periods, 56 or 64 years later, they are shifted by about a month, so they are again in step with the moon, plus/minus two days. For example, the Venus data fit well with a beginning of Hammurabi's reign in 1848 BC (the so-called "long" chronology), but also with 1792, 1784 and 1728 BC (the two "middle" and the "short" chronologies). These are the four most popular chronologies among historians. Around 1980, I came back to the problem and showed that the astronomical evidence overwhelmingly favored the long chronology, and that the middle and short ones in all likelihood were incorrect. This was based on a relatively delicate statistical argument, combining robust, frequentist and Bayesian methods. Among Assyriologists, some were convinced and some were not. Some distrusted the corrupt data, and some rejected the long chronology because it leaves a dark period in the middle of the second millennium, a hole without historical information. There is still a big discussion which chronology is correct.

**AB:** In 1980, what made you go back to this problem again? You thought you had some additional insights or computing was better?

**PH:** Earlier I had used only the Venus data and I wanted to use the month-length data as well. The month-length data had been used already by the first book on the Venus data in the 1920s. But now we had much more data and I thought we could do better. So this was the reason I went back. I was invited to present this material at an Assyriology meeting in London in 1982, which I did. The next effort was when I retired and had more time on hand.

**HK:** What led to this book in front of us? [Pointing to a book on the table.[10]]

**PH:** That book had a very long gestation period, dating back to the time when I helped van der Waerden write his book on Babylonian astronomy. In that context I had to look at the lunar eclipse observation texts which had just been published in 1955. I tried to figure out what they really said. I did that and produced a handwritten manuscript as a basis for remarks in van der Waerden's book. Later, in 1973, I broke a leg and was immobilized for a while, so I went over the material once more and wrote a more complete manuscript, in English this time. Some people referred to this manuscript as "Huber's samizdat." Somebody claimed it was the most quoted unpublished manuscript in the area, which doesn't mean much, because not too many people work in that area. From time to time people asked for a copy. It was Salvo De Meis who pushed me into publishing. This was in the late 1990s.

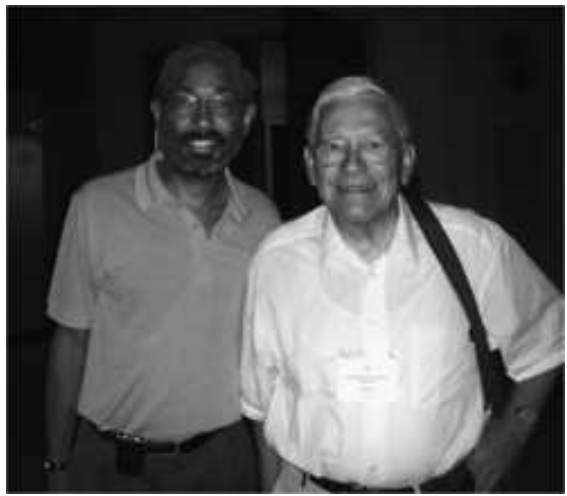

FIG. 9. *Houston, 2007. Third Erich L. Lehmann Symposium. Chance encounter with a former Ph.D. Student: Emery Brown, Professor of Computational Neurology, MIT/Harvard.*

**AB:** ...and De Meis is...?

**PH:** ...a nuclear engineer by profession, but he also does ancient astronomy as a hobby. We got together and expanded the manuscript. Since the former draft, other texts had been published, so we included all of them. We did some data analysis to solve the problem of the difference between civil time and dynamical time. Recall that civil, universal time is essentially Greenwich time, based on the rotation of the earth; ephemeris or dynamical time is based on the uniform time scale underlying the gravitational theories.

**AB:** Do you see any other projects along these lines, or open problems?

**PH:** There are many problems, perhaps not enough data. Right now I am looking into something very nonstatistical, namely, the old Babylonian understanding of Sumerian grammar. Old Babylonian grammar texts are fascinating because they are the earliest serious grammatical documents, roughly from Hammurabi's time. Akkadian, spoken by the Babylonians, is a Semitic language; Sumerian is an agglutinating language without links to any other known language. Sumerian died out approximately 2000 BC as a spoken language, but it continued to be used until the very end of cuneiform, that is, until about year 0. Just as with the old Babylonian mathematical texts, the interesting thing about these texts is the history of science aspect. They offer extensive, disciplined verbal paradigms with the Sumerian forms on the left-hand side and the corresponding Akkadian forms on the right-hand side. Hence it should be possible to extract the old Babylonian understanding of Sumerian verbal grammar from these texts. This looks like an interesting challenge to me.[11]

---

[10]HUBER, P. J. and DE MEIS, S. (2004). *Babylonian Eclipse Observations from 750 BC to 1 BC.* Milan: IsIAO-Mimesis, 2004.

**HK:** Peter, thank you for these fascinating stories. We wish you success with your adventures in Assyriology and hope you will keep in touch with statistics.

---

[11] HUBER, P. J. (2007). On the Old Babylonian Understanding of Grammar: A Reexamination of OBGT VI-X. *J. Cuneiform Studies* **59** 1–17.